\documentstyle[prl,aps,epsfig]{revtex}         

\begin{document}
%
%
%
\twocolumn[\hsize\textwidth\columnwidth\hsize\csname
@twocolumnfalse\endcsname

\title{Enhancement and suppression of spontaneous emission and light scattering by quantum degeneracy}
\author{A. G\"{o}rlitz, A. P. Chikkatur, and W. Ketterle }
\address{Department of Physics and Research Laboratory of
Electronics, \\ Massachusetts Institute of Technology, Cambridge, MA 02139}
\date{\today}
\maketitle
\begin{abstract}
Quantum degeneracy modifies light scattering and spontaneous emission. For fermions, Pauli blocking
leads to a suppression of both processes. In contrast, in a weakly interacting Bose-Einstein
condensate, we find spontaneous emission to be enhanced, while light scattering is suppressed. This
difference is attributed to many-body effects and quantum interference in a Bose-Einstein
condensate.
\end{abstract}
 \vskip1pc ]

For a long time, spontaneous emission and light scattering were
regarded as intrinsic properties of atoms.  However, quantum
electrodynamics (QED) revealed the connection between these
phenomena and the electromagnetic modes of the vacuum. Spontaneous
emission and scattering can only take place when a vacuum mode is
available to accommodate the emitted or scattered photon. The
insight that small cavities can be used to modify the vacuum is
exploited in cavity-QED experiments \cite{ber94}. Recent
breakthroughs in the experimental realization of gaseous quantum
degenerate systems of bosons
\cite{ande95,davi95bec,brad97bec,frie98} and fermions
\cite{dema99} have opened the possibility to use quantum
degeneracy to modify the emission and scattering behavior of
atoms. While a cavity changes the mode structure for the scattered
or emitted photon, the presence of a quantum degenerate system
modifies the modes for the recoiling atom.

Quantum degeneracy influences spontaneous emission and light scattering  via  the presence of
population in the final quantum states of the process. Generally, transition rates between an
initial state with population $N_1$ and a final state with population $N_2$ are proportional to
$N_1 (1+N_2)$ for bosons and to $N_1 (1 - N_2)$ for fermions. This reflects the well-known fact
that in a bosonic system the transition into an already occupied state is enhanced by bosonic
stimulation, while in fermionic systems, occupation of a state prevents a transition into this
state by Pauli blocking.

In this note, we show theoretically that in a weakly interacting
Bose-Einstein condensate spontaneous emission is enhanced and
calculate the enhancement factor.  We compare this result to
light scattering in a Bose-Einstein condensate (BEC), which is
suppressed as we have shown experimentally and theoretically in
previous work \cite{stam99phon}. In contrast, in fermionic
systems quantum degeneracy leads to a suppression of \emph{both}
spontaneous emission and light scattering
\cite{ruos99,dema98,helm90}.

For simplicity we consider a homogeneous system of atoms at zero
temperature, which exhibits all the significant features. The
system is assumed to consist of $N$ atoms with mass $m$ at a
density $n$ in a volume $V$, which is sufficiently large such that
the summations over quantum states can be approximated by
integrals.

A non-interacting BEC at $T=0$ can be described by a single-particle ground-state wavefunction with
amplitude $\sqrt{N}$. Thus, the occupation of quantum states with wavevector $\bf{k}$ is given by
$N(\bf k \mathnormal ) = N\, \delta(\bf k \mathnormal )$. Since the final states for spontaneous
emission and  light scattering with a finite momentum transfer $\bf q $ are not occupied, both
processes occur at the single-atom rate.

This situation changes drastically for an interacting condensate.
Here, two atoms in the zero-momentum state are coupled to states
with momenta $+{\bf k}$ and $-{\bf k}$. This changes the
excitation spectrum $\omega(k)$ from the free-particle form,
$\hbar \omega(k) =E_r(k)$, to the spectrum of Bogoliubov
quasi--particles, $\hbar \omega(k) = \sqrt{E_r(k) (E_r(k) + 2
\mu)}$. Here, $\mu$ is the chemical potential and $E_r(k) =
\hbar^2 k^2/2 m$ the recoil energy associated with the momentum
$k$. The chemical potential is a measure for the interatomic
interactions and is related to the $s$-wave scattering length $a$
by $\mu= 4 \pi \hbar^2 a n/m$.

The atom-atom interaction admixes pair correlations into the
ground state wavefunction $|\text{BEC},N \rangle$ of a BEC with
$N$ atoms, yielding the structure \cite{huan87}
\begin{eqnarray}
\label{equ:bec_ground_state}
 |\text{BEC},N \rangle &=& |N,\,0,\,0\rangle \\ \nonumber &&-
\alpha |N-2,\, 1,\,1\rangle + \alpha^2 |N-4,\, 2,\, 2 \rangle + \dots ,
\end{eqnarray}
where $\alpha=1-1/u_k^2$. Here $|N_0, N_k, N_{-k}\rangle$ denotes
a state with $N_0$ atoms in the zero-momentum state and $N_{\pm
k}$ atoms in states with momentum $\pm k$. In
Eq.(\ref{equ:bec_ground_state}) a summation over all momenta $\bf
k$ is implicitly assumed. The average population of momentum
states is given by $N(k)= u_{k}^{2}-1 = v_{k}^{2}$, where $u_k =
\cosh \phi_k$, $v_k = \sinh \phi_k$ and $\tanh 2\phi_k = \mu /
(E_r(k) + \mu)$.

To study the effect of the presence of a BEC on spontaneous
emission, we consider an excited atom at rest added to a BEC of
$N$ ground-state atoms. This system is described by an initial
state $|i\rangle = \hat{a}_{e,0}^\dagger |\text{BEC},N \rangle$,
where $\hat{a}_{e,0}^\dagger$ creates an electronically excited
atom at rest. We use Fermi's golden rule to obtain the rate for
spontaneous emission.  The only difference to the single-atom
spontaneous decay rate $\Gamma$ comes from the overlap matrix
elements to the final momentum state $|f\rangle$, $\langle
f|\hat{a}_{k_L}^\dagger \hat{a}_{e,0} | i \rangle$ where
$\hat{a}_{k_L}^\dagger$ is the creation operator for a free
ground state atom with momentum $k_L$.  Summing over all final
states one arrives at
\begin{eqnarray}
\label{equ:spont_BEC} \gamma_{\text{BEC}}=\Gamma \,   \langle
\text{BEC},N |\hat{a}_{k_L} \hat{a}_{k_L}^\dagger|\text{BEC},N
\rangle \, .
\end{eqnarray}
Thus, the spontaneous emission rate is proportional to the square
of the norm of the state vector $|e^+ \rangle =
\hat{a}_{k_L}^\dagger|\text{BEC},N\rangle $.

To calculate the norm of $|e^+ \rangle$ explicitly, we transform
to Bogoliubov operators by substituting $\hat{a}_k = u_k \hat{b}_k
- v_k \hat{b}^\dagger_{-k}$. The operators $\hat{b}^\dagger_{k}$
and $\hat{b}_{k}$ are  the creation and annihilation operators for
the  microscopic quasi-particle excitations of a weakly
interacting condensate. Hence, the many-body ground-state
wavefunction of the condensate $|\text{BEC},N \rangle$
corresponds to the quasi-particle vacuum defined by the relation
$\hat{b}_k |\text{BEC},N\rangle \equiv 0, \forall k$. Thus, we
obtain
\begin{eqnarray}
\label{equ:f_bose_spont}
 F_{\text{Bose}}^{\text{spont}} &=& \langle e^+|e^+ \rangle = u_{k_L}^2
 =  1+ N(k_L)  \\ \nonumber &=& \left( \cosh \,\left( \frac{1}{2}\, \tanh^{-1}\left(\frac{k_s^2}{k_L^2/2 +
 k_s^2}\right)\right) \right)^2 \, ,
\end{eqnarray}
where $\hbar k_s = m c$ is the momentum of an atom moving at the
speed of sound, which is related to the chemical potential by
$\mu= mc^2$. Enhancement of spontaneous emission in a BEC is
significant if $k_s$ becomes comparable to the wavevector $k_L$ of
the emitted photon since for small momentum transfer $u_{k_L}^{2}=
k_s^2/k_L^2$.  Eq.(\ref{equ:f_bose_spont}) and its interpretation
are the major results of this paper. It should be noted, that this
enhancement of spontaneous emission in a BEC is distinctly
different from the phenomenon of superradiance as discussed by
Dicke \cite{dick54}. Supperadiance occurs due to the \em
collective \em emission of radiation in a sample of atoms prepared
in the excited state. In contrast, the enhancement discussed here
affects a \em single \em electronically excited atom due to
macroscopic occupation in the final state of the recoiling atom.

The result of Eq.(\ref{equ:f_bose_spont}) that spontaneous
emission in a weakly interacting BEC is enhanced, is striking in
view of our earlier finding \cite{stam99phon} that light
scattering in a BEC is suppressed. The operator describing a light
scattering event with momentum transfer $\bf{q}$ is the Fourier
transform of the atomic density operator $\hat{\rho} (\bf q
\mathversion{normal} ) $$ = \sum_m \hat{a}^\dagger_{m+q}
\hat{a}_m$. If $\hat{\rho}(\bf q \mathnormal)$ acts on
$|\text{BEC},N\rangle$, only terms involving the zero-momentum
state $m=0$ yield significant contributions. By applying Fermi's
golden rule we found that the scattering rate is proportional to
the norm of the state vector
\begin{eqnarray}
  |e\rangle & \approx & \frac{(\hat{a}^\dagger_q \hat{a}_0 + \hat{a}^\dagger_0
  \hat{a}_{-q}) |\text{BEC},N \rangle}{\sqrt{N}} \\ \nonumber
  & \approx & (\hat{a}^\dagger_q  + \hat{a}_{-q}) |\text{BEC},N \rangle \\ \nonumber
 & = & |e^+\rangle + |e^-\rangle\, ,
\end{eqnarray}
where we have replaced $\hat{a}^\dagger_0$ and $\hat{a}_0$ by $\sqrt{N}$ following the usual
Bogoliubov formalism \cite{bogo47}. After transforming to Bogoliubov operators we obtain a
suppression factor of
\begin{equation}
 S_{\rm Bose}(q) = \langle e | e \rangle = (u_q\, -\, v_q)^2 \, ,
 \label{equ:norm_e}
\end{equation}
 which is the static structure factor for a BEC.
 Generally, the static structure factor is the normalized
response of a system  to a  perturbation with wavevector $\bf q$. For small $\bf q$, corresponding
to  phonon-like quasi-particle excitations, $S_{\rm Bose}(q)= \hbar q/2 m c $ approaches zero.
Light with wavevector $k_L$ scattered at an angle $\theta$ imparts a momentum $\hbar q = 2 \hbar
k_L \sin(\theta/2)$ to the atomic system. By integrating Eq.(\ref{equ:norm_e}) over all possible
scattering angles $\theta$ and accounting for the dipolar emission pattern, we find that Rayleigh
scattering from a BEC is suppressed by a factor \cite{stam99phon}
\begin{eqnarray}
\label{equ:f_bose_scatt} &&F_{\text{Bose}}^{\text{scatt}} =
\frac{k_s}{\sqrt{k_s^2+k_L^2}} \left(\frac{15}{8}
\,\frac{k_s^5}{k_L^5} \, +  \, \frac{23}{8} \, \frac{k_s^3}{k_L^3}
\, + \, 2 \, \frac{k_s}{k_L} \, + \, \frac{k_L}{k_s} \right)
\\ \nonumber && \ \ - \, \left(\frac{15}{8} \, \frac{k_s^6}{k_L^6} \, + \,
\frac{9}{4} \, \frac{k_s^4}{k_L^4} \, + \, \frac{3}{2} \,
\frac{k_s^2}{k_L^2} \right) \, \tanh^{-1} \left(
\frac{k_L}{\sqrt{k_s^2+k_L^2}} \right) \ .
\\ \nonumber
\end{eqnarray}

For comparison, we briefly summarize the suppression of
spontaneous emission and light scattering for a fermionic system.
A Fermi gas at $T = 0$ with Fermi momentum $\hbar k_F$ is
characterized by $N({\bf k}) = \theta(k_F-k)$, i.e. all momentum
states with $k < k_F= (6 \pi n)^{1/3}$ are occupied. If we add an
electronically excited atom at rest to the Fermi sea, its
spontaneous decay rate is suppressed by a factor
\begin{equation}
F^{\text{spont}}_{\text{Fermi}} = 1-N(k_{L})=\theta(k_{L}-k_F)\, .
\label{equ:f_fermi_spont}
\end{equation}

When off-resonant light with initial wavevector ${\bf k_L}$ is
scattered from a filled Fermi sphere into an outgoing wave with
final wavevector ${\bf k_L}+{\bf q}$, the scattering rate is
suppressed by \cite{pine88one}
\begin{eqnarray}
\label{equ:fermi_structure_factor} S_{\text{fermi}}({\bf q}) &=&
\int {\rm d} {\bf k}\ N({\bf k}) (1-N({\bf k+q}))\\ \nonumber &=&
\cases {\frac{3 q}{4 k_F} - \frac{q^3}{16 k_F} & \text{if $0 < q
<2 k_F$}, \cr\cr 1 & \text{if $q >2 k_F$}.}
\end{eqnarray}
Eq.(\ref{equ:fermi_structure_factor}) is the static structure
factor for a Fermi gas at zero temperature. Integrating over all
possible scattering angles $\theta$ and accounting for the dipolar
emission pattern, we find that the total suppression factor for
Rayleigh scattering from a Fermi sea is given by
\begin{equation}
\label{equ:f_fermi_scatt} F^{\text{scatt}}_{\text{Fermi}} = \cases
{\frac{69}{70} \frac{k_L}{k_F} - \frac{43}{210}
\frac{k_L^3}{k_F^3} & \text{if $k_L <k_F$} \cr\cr 1 - \frac{3}{10}
\frac{k_F^2}{k_L^2} \, + \, \frac{9}{70} \frac{k_F^4}{k_L^4}  -
\frac{1}{21} \frac{k_F^6}{k_L^6} & \text{if $k_L>k_F$}.}
\end{equation}

\begin{figure}
\epsfig{file=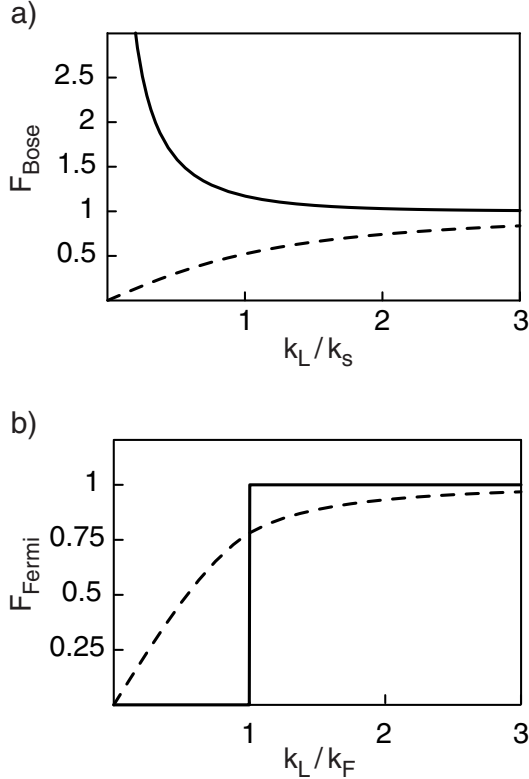,width=7cm} \caption{Modification of spontaneous
emission (solid line) and light scattering (dashed line) due to
quantum degeneracy. In a) we have plotted the enhancement factor
for spontaneous emission and the suppression factor for light
scattering for a weakly interacting Bose-Einstein condensate as a
function of the light wavevector $k_L$ in units of $k_s$, the
wavevector of an atom moving at the speed of sound. In b) the
suppression factors for spontaneous emission and light scattering
in a  Fermi gas at $T=0$ are plotted as a function of $k_L$ in
units of the Fermi wavevector $k_F$.} \label{fig1}
\end{figure}

Fig.\,\ref{fig1} shows the influence of quantum degeneracy on the atom-light interaction. Using
Eqs.(\ref{equ:f_bose_spont}),(\ref{equ:f_bose_scatt}),(\ref{equ:f_fermi_spont}), and
(\ref{equ:f_fermi_scatt}) we have plotted the rates for spontaneous emission (solid lines) and
light scattering (dashed lines), normalized by the single-atom rates, for a weakly interacting BEC
(Fig.\,\ref{fig1}a) and a degenerate Fermi gas (Fig.\,\ref{fig1}b). A significant deviation from
the free-particle rate is clearly observable if the photon-momentum is comparable to $k_s$ for
bosons and $k_F$ for fermions.

Let us now discuss the intrinsic difference between atom-light interaction in a BEC and in a
degenerate Fermi gas. In a Fermi sea, the suppression of both light scattering and spontaneous
emission is a \emph{single-particle} effect caused by the non-availability of final states due to
the Pauli exclusion principle.  Indeed, one would obtain the same result for spontaneous emission
as for light scattering (Eq.(\ref{equ:f_fermi_scatt})) if the initial momentum of the excited atom
were randomly distributed over the Fermi sphere.

In an interacting BEC the situation is significantly different
because pair correlations in the ground state, i.e.
\emph{many-body} effects, are responsible for both the enhancement
of spontaneous emission and suppression of light scattering. The
finite population in states with $k \neq 0$ due to quantum
depletion lends a very intuitive explanation for the enhancement
of spontaneous emission. If the interactions in the condensate are
sufficiently strong such that momentum states with $k=k_{L}$ have
non-negligible occupation, spontaneous emission of an atom at rest
is enhanced by bosonic stimulation. This intuitive argument is
correct, but it would incorrectly predict that light scattering is
also enhanced.

The suppression of light scattering occurs due to the correlation
between the admixtures of states with momentum $k$ and $-k$. This
leads to a destructive quantum interference between the two
processes $|N,0,0\rangle + \hbar q \rightarrow |N,1,0 \rangle$ and
$|N-2,1,1\rangle + \hbar q \rightarrow |N,1,0 \rangle$, in which
either an excitation with momentum $\bf{-q}$ is annihilated or an
excitation with momentum $\bf{q}$ is created. Both processes
transfer momentum $\bf{q}$ to the condensate and are individually
enhanced by bosonic stimulation.  Therefore, a simple rate
equation model would predict enhanced light scattering. However,
since the initial states are correlated the two processes leading
to the same final state interfere destructively for a BEC with
repulsive interactions and light scattering is suppressed.

The results presented above for suppressed light scattering in both bosonic and fermionic systems
also apply to the scattering of massive impurities. This has been studied theoretically for
degenerate Bose \cite{timm98} and Fermi \cite{ferr99,holl00} gases, and was experimentally observed
in a BEC \cite{chik00}.

How strong would the enhancement of spontaneous emission in
currently realized Bose-Einstein condensates be? Condensates of
$^{23}\text{Na}$ atoms confined in an optical trap  have reached a
density of $3 \times 10^{15} \, {\rm cm}^{-3}$ \cite{stam98odt}.
For this density the speed of sound $\hbar k_s/m = 2.8 \rm \,
cm/s$ and the recoil velocity $\hbar k_L/m = 2.9 \rm \, cm/s$ are
approximately equal and we find $N(k_L) \approx 0.15$. Thus, the
observation of enhanced spontaneous emission in a BEC is within
experimental reach. Excited atoms at rest could be produced by
Doppler free two-photon excitation, a scheme already used to
probe condensates of atomic hydrogen using the $1s \rightarrow
2s$ transition \cite{frie98}. Another possibility is to inject
ground-state atoms with momentum $\hbar k_L$ into a condensate
and use a counter-propagating laser beam to excite them and bring
them to rest.  The enhancement of spontaneous emission could then
be observed as frequency broadening of the absorption line.

The fact that light scattering is suppressed, but spontaneous
emission is enhanced, could be exploited for studies of
decoherence in a BEC.  When a photon is absorbed by a BEC (the
first step of light scattering), it creates a (virtual) excited
state that has an external wavefunction which includes pair
correlations.  Any decoherence of this coherent superposition
state, for example by interaction with the thermal cloud, could
destroy the interference effect discussed above and turn the
suppression of light scattering into an enhancement.

In conclusion, we have discussed suppression and enhancement of
light scattering and spontaneous emission in quantum degenerate
systems, and shown that in a weakly interacting BEC, the quantum
depletion can enhance spontaneous emission by bosonic
stimulation. This contrasts earlier results on suppressed light
scattering in a BEC. As we have shown, both the reduced light
scattering and the enhanced spontaneous emission in a BEC are
related to quantum depletion of the condensate.  However, the
enhanced spontaneous emission appears to be physics beyond the
Gross-Pitaevskii equation, while the static structure factor S(q)
and the reduced light scattering can be obtained from the
Gross-Pitaevskii equation \cite{kett00lho}.

During preparation of this manuscript we found a theoretical paper
\cite{maze00} which predicts an enhancement of light scattering in
a weakly interacting Bose-Einstein condensate contrary to our
findings.

The authors acknowledge valuable discussions with B. E. Stoicheff
and D. M. Stamper-Kurn and critical reading of the manuscript by
S. Inouye. This work was supported by NSF, ONR, ARO, NASA, and
the David and Lucile Packard Foundation. A.P.C. would like to
acknowledge additional support from the NSF.

\end{document}